\documentclass[12pt]{article}
\newcommand{\eq}{\begin{equation}}
\newcommand{\feq}{\end{equation}}
\newcommand{\eqn}{\begin{eqnarray}}
\newcommand{\feqn}{\end{eqnarray}}
\newcommand{\arr}{\begin{eqnarray*}}
\newcommand{\farr}{\end{eqnarray*}}

\def\la{\lambda}

\def\si{\sigma}
\def\t{\tau}
\def\om{\omega}

\def\lb{\label}
\begin{document}
\begin{titlepage}
\begin{flushright}
INFNCA-TH0014 \\
IFUM 661/FT \\
hep-th/0009185
\end{flushright}
\vspace{.3cm}
\begin{center}
\renewcommand{\thefootnote}{\fnsymbol{footnote}}
{\Large \bf AdS$_2$ Gravity as Conformally Invariant Mechanical System}
\vfill%\vskip 15mm%27.mm
{\large \bf {M.~Cadoni$^{1,a}$\footnote{email: cadoni@ca.infn.it},
P.~Carta$^{1,a}$\footnote{email: carta@ca.infn.it},
D.~Klemm$^2$\footnote{email: dietmar.klemm@mi.infn.it},
and S.~Mignemi$^{3,a}$\footnote{email: mignemi@ca.infn.it}}}\\
\renewcommand{\thefootnote}{\arabic{footnote}}
\setcounter{footnote}{0}
\vfill%\vskip 7mm%1cm
{\small
$^1$ Universit\`a di Cagliari, Dipartimento di Fisica,\\
Cittadella Universitaria, 09042 Monserrato, Italy.\\
\vspace*{0.4cm}
$^2$ Universit\`a degli Studi di Milano, Dipartimento di Fisica and\\
INFN, Sezione di Milano,
Via Celoria 16,
20133 Milano, Italy.\\
\vspace*{0.4cm}
$^3$ Universit\`a di Cagliari, Dipartimento di Matematica,\\
Viale Merello 92, 09123 Cagliari, Italy.\\
\vspace*{0.4cm}
$^a$ INFN, Sezione di Cagliari.}
\end{center}
\vfill
\begin{center}
{\bf Abstract}
\end{center}
{\small
We show that two-dimensional (2D) AdS gravity
induces on the spacetime boundary a conformally invariant dynamics
that can be described in terms of a
de Alfaro-Fubini-Furlan model coupled
to an external source with conformal dimension two.
The external source encodes the information about the gauge
symmetries of the 2D gravity system.
Alternatively, there exists a description in terms of a
mechanical system with anholonomic constraints.
The considered systems are invariant under the action of the 
conformal group generated by  a Virasoro
algebra, which occurs also as asymptotic symmetry algebra of two-dimensional
anti-de~Sitter space. We calculate the central charge
of the algebra and find perfect
agreement between statistical and thermodynamical entropy of AdS$_2$
black holes.}

\end{titlepage}

Recent investigations have brought evidence of a deep
connection between
two-di\-mens\-ion\-al (2D) gravitating systems and conformal 
mechanics \cite{claus}.
The most natural context to test this conjecture is the
Anti-de~Sitter/conformal field theory (AdS/CFT) correspondence \cite{adscft}
in two spacetime dimensions \cite{strom}.
In fact, for $D=2$ this correspondence essentially states that
gravity on AdS$_2$ should be described by a conformally
invariant quantum mechanics.

Most of the progress in this direction has been achieved in
the context of 2D dilaton gravity, mainly because of
the simplicity of the model \cite{CM99,CM00,navarro}.
For these models the conformal symmetry
(generated by a Virasoro algebra) has a natural interpretation
in terms of the asymptotic symmetries of the gravitational system.
Moreover, the central charge of the algebra,
whose value is crucial for calculating the
statistical entropy of 2D black holes, can be calculated using the
deformation algebra of the boundary of AdS$_2$ \cite{CM99}.

Despite the simplicity of the model, the various attempts to
identify the conformal quantum mechanics that should be dual to
gravity on AdS$_2$, and to calculate the entropy of 2D black holes
by counting states of the CFT, met only partial success 
\cite{CM99,CM00,navarro}.
The conformal mechanics involved could not be identified and a
mismatch of a $\sqrt 2$ factor between thermodynamical and
statistical entropy was found.

The puzzle has become even more intricate in view of the results
of Ref.~\cite{CC}. There it was shown that in $D=2$ the AdS/CFT
correspondence has a realization in terms of a two-dimensional CFT, which
is essentially a theory of open strings with
Dirichlet boundary conditions. Moreover, it was shown that the
degeneracy of
states of the 2D CFT explains correctly the thermodynamical
entropy of 2D black holes.

In this letter we show that the boundary dynamics
induced by AdS$_2$ dilaton gravity can be described by a
De Alfaro-Fubini-Furlan (DFF) model \cite{dff} coupled
to an external source with conformal dimension two.
The external source encodes the information about the gauge
symmetries of the 2D gravity system.
Alternatively, the dynamics of the boundary fields
admits an equivalent description in terms of a
mechanical system with anholonomic constraints.
In both cases, the mechanical system is invariant under the action
of the full one-dimensional conformal group generated by a
Virasoro algebra, which also appears as
asymptotic symmetry algebra of AdS$_2$.
We compute the central charge of this algebra and
find perfect
agreement between statistical and thermodynamical entropy of AdS$_2$
black holes.

Our starting point is the Jackiw-Teitelboim (JT) model \cite{jackiw},
with action
\eq
I = \frac{1}{2}\int d^2x \sqrt{-g}\eta[R + 2\lambda^2],
      \label{JT}
\feq
where $\eta$ represents the dilaton.
Two-dimensional anti-de~Sitter space, or more
generally black holes in AdS$_2$, are solutions of this model
\cite{CM95},
\eq\lb{bh}
ds^2 = -(\lambda^2r^2-a^2)dt^2 + (\lambda^2r^2-a^2)^{-1}dx^2,\quad
\eta=\eta_{0}\la x,
\feq
where $a$ is given in terms of the black hole mass $M$, $a^{2}=2M/\la$.
The thermodynamical black hole entropy is given by
\eq\lb{entropy}
S=4\pi \sqrt{ \eta_{0}M\over 2\la}.
\feq
The asymptotic symmetries of AdS$_2$
are defined as the transformations which leave the asymptotic form
of the metric invariant, i.~e.~they preserve the large $r$ behavior
\eqn
g_{tt} &=& -\lambda^2r^2 + \gamma_{tt}(t) + {\cal O}\left(\frac{1}{r^2}\right),
           \nonumber \\
g_{tr} &=& \frac{\gamma_{tr}(t)}{\lambda^3r^3} + {\cal O}\left(\frac{1}{r^5}
           \right), \nonumber \\
g_{rr} &=& \frac{1}{\lambda^2r^2} + \frac{\gamma_{rr}(t)}{\lambda^4r^4}
           + {\cal O}\left(\frac{1}{r^6}\right),
\label{d1}
\feqn
where the fields $\gamma_{\mu\nu}$ parametrize the
first sub-leading terms in the
expansion and can be interpreted as deformations of the boundary.

The asymptotic form (\ref{d1}) is preserved by infinitesimal
diffeomorphisms $\chi^{\mu}(x,t)$ of the form \cite{CM99}
\eqn
\chi^t &=& \epsilon(t) + \frac{\ddot{\epsilon}(t)}{2\lambda^4r^2} +
           \frac{\alpha^t(t)}{r^4} + {\cal O}\left(\frac{1}{r^5}\right),
           \nonumber \\
\chi^r &=& -r\dot{\epsilon}(t) + \frac{\alpha^r(t)}{r} +
           {\cal O}\left(\frac{1}{r^2}\right), \label{d2}
\feqn
where $\epsilon(t)$ and $\alpha^{\mu}(t)$ are arbitrary functions,
the $\alpha^{\mu}$ describing pure gauge diffeomorphisms.
In Ref.~\cite{CM99} it was shown that the symmetries (\ref{d2}) generate
a Virasoro algebra.

The asymptotic behavior of the scalar field $\eta$, compatible
with the transformations (\ref{d2}), must take the form
\eq
\eta = \eta_0\left(\lambda\rho(t)r + \frac{\gamma_{\eta}(t)}{2\lambda r}\right)
       + {\cal O}\left(\frac{1}{r^3}\right), \label{d3}
\feq
where $\rho$ and $\gamma_{\eta}$ play a role analogous to that of the
$\gamma_{\mu\nu}$. Introducing the new fields, invariant under the pure
gauge diffeomorphisms parametrized by $\alpha^\mu$,
\eqn
\beta = \frac 12 \rho \gamma_{rr} + \gamma_{\eta}, \nonumber \\
\gamma = \gamma_{tt} - \frac 12 \gamma_{rr},
\feqn
the equations of motion following from
the action (\ref{JT})
yield in the limit $r\to\infty$\footnote{At first sight, it seems not
necessary to require (\ref{d5a}), since it comes from the leading term
in the stress tensor component $T_{rt}$, which is of order $1/x^2$ \cite{CV}.
However, $T_{\mu\nu}$, transforming as a tensor,
is clearly not invariant under coordinate transformations. In fact,
in the light-cone coordinates used below, (\ref{d5a}) originates from
an order $1$ term, so requiring (\ref{d5a}) is really necessary for
consistency.}
\eq
\lambda^{-2}\ddot{\rho} = \rho\gamma - \beta, \label{d5}
\feq
\eq
\dot{\rho}\gamma + \dot{\beta} = 0. \label{d5a}
\feq
Eqs. (\ref{d5}) and (\ref{d5a}) determine a mechanical system with
anholonomic constraint, since the one-form
\eq
\omega \equiv \gamma d\rho + d\beta
\feq
is not exact. The Lagrange equations of the first kind
for the fields $\varphi_i = \{\rho,\beta,\gamma\}$ read
\eq
F_i - m_i\ddot{\varphi}_i + \Lambda \omega_i = 0, \label{lagr}
\feq
where $F_i$ is the force that can be derived from a potential $U$,
$F_i = -\partial_i U$, $m_i$ denote the masses of the fields, $\Lambda$
is a Lagrange multiplier, and the $\omega_i$ are the components of the
one-form $\omega$. If we choose
\eq
m_{\rho} = \lambda^{-1}, \quad m_{\beta} = m_{\gamma} = 0,
\feq
and
\eq
U = \lambda\beta\rho,
\feq
the Lagrange equations (\ref{lagr}) yield (\ref{d5}), together with
the Lagrange multiplier $\Lambda = \lambda\rho$. Before we proceed, we
note that from Eqs. (\ref{d5}) and (\ref{d5a}), one gets the conservation law
\eq\label{energy}
T + U = \frac 12 \lambda^{-1}\dot{\rho}^2 + \lambda\beta\rho
= {\mathrm{const.}}
\feq
Notice that $T+U$ is essentially the mass of the black holes considered in
\cite{CM99}.\\
The boundary fields $\varphi_i$ span a representation of the full
infinite dimensional group
generated by the Killing vectors (\ref{d2}).
In fact, under the asymptotic symmetries (\ref{d2}), they
transform as
\eqn
\delta \rho &=& \epsilon\dot{\rho} - \dot{\epsilon}\rho, \nonumber \\
\delta \beta &=& \epsilon\dot{\beta} + \dot{\epsilon}\beta +
                 \frac{\ddot{\epsilon}\dot{\rho}}{\lambda^2}, \nonumber \\
\delta \gamma &=& \epsilon\dot{\gamma} + 2\dot{\epsilon}\gamma -
                  {\stackrel\dots\epsilon\over \lambda^2}.
\label{d6}
\feqn
The above transformations are easily recognized as (anomalous)
transformation laws for conformal fields of weights
$-1,1,2$ respectively. We are interested in the transformation laws
of the equation of motion (\ref{d5}), the constraint (\ref{d5a}), and
the conserved charge (\ref{energy}).
Using (\ref{d6}), we get
\eqn\label{d7}
\lefteqn{\delta[\gamma\dot{\rho} + \dot{\beta}] = } \nonumber \\
& & \epsilon\frac{d}{dt}[\gamma\dot{\rho} + \dot{\beta}] +
2\dot{\epsilon}[\gamma\dot{\rho} + \dot{\beta}] +
\ddot{\epsilon}[-\gamma\rho + \beta + \frac{\ddot{\rho}}{\lambda^2}],
\feqn
\eqn\label{d7a}
\lefteqn{\delta[-\gamma\rho + \beta +
\frac{\ddot{\rho}}{\lambda^2}] = } \nonumber \\
& & \epsilon\frac{d}{dt}[-\gamma\rho + \beta +
\frac{\ddot{\rho}}{\lambda^2}] + \dot{\epsilon}
[-\gamma\rho + \beta +
\frac{\ddot{\rho}}{\lambda^2}],
\feqn
\eq
\delta[T+U] = \epsilon\frac{d}{dt}[T+U].
\feq
We see that the constraint transforms like a conformal field of weight 2
with anomaly term. The conformal weights of the equation of motion and
the conserved charge are 1 and 0 respectively, and anomalies are absent.
The above equations imply that on-shell the constraint and the equation
of motion are invariant under the transformations (\ref{d6}).\\

Alternatively, we may describe the dynamical system (\ref{d5})
in terms of the DFF model \cite{dff} of conformal mechanics,
coupled to an external source.
To this aim, we start from the conservation law (\ref{energy}), i.~e.~$T+U=c$.
Introducing the new field $q=\sqrt{\rho/\lambda}$, which has conformal
dimension $-1/2$, and eliminating $\beta$ from (\ref{energy}) by means
of (\ref{d5}), we arrive at the equation
\eq
\ddot{q} - \frac{g}{q^3} = \frac{\lambda^2}{2}\gamma q, \label{dffeq}
\feq
with $g=-c/(2\lambda)$, whereas from Eq.~(\ref{energy}) follows
\eq
\frac{\dot{q}^2}{2} + \frac{g}{2q^2} = -\frac{\lambda^2}{4}\beta.
\label{dffham}
\feq
One can easily check the equivalence of the system (\ref{d5}),
(\ref{d5a}) with (\ref{dffeq}), (\ref{dffham}).
Moreover, (\ref{dffeq}) is easily recognized as the equation
of motion for the DFF model coupled to an external source $\gamma$.

At this point it is straightforward to write down an action for the dynamical
system (\ref{dffeq}). It is given by
\eq
I = \int dt \left[\frac 12 \dot{q}^2 - \frac{g}{2q^2} + \frac 14 \lambda^2
    \gamma q^2\right]. \label{1daction}
\feq
(\ref{1daction}) resembles very much the IR-regularized action proposed
by DFF \cite{dff}, the only difference consists in the fact that the
external source $\gamma$, which couples to the field $q$, is not constant,
but represents an operator of conformal dimension two.
Note that in the calculation of $\delta I$, $\gamma$, being an external source,
is not varied.
One can easily show that the action is (up to a total derivative)
invariant under the conformal transformations (\ref{d6}).

The dynamical system described by Eqns.~(\ref{d5}), (\ref{d5a}) (or
equivalently by (\ref{dffeq})), defines a one-dimensional
conformal field theory (CFT$_1$). The invariance group of the model
coincides with the group of asymptotic symmetries of AdS$_2$, and can
be realized as the diff$_1$ group describing time-reparametrizations
$\delta t= \epsilon(t)$.
In analogy with 2D CFT, one would like to identify the
stress-energy tensor $T_{tt}$ associated with the CFT$_{1}$.
This analogy suggests that $T_{tt}$ is proportional to the constraints
(\ref{d5a}),
\eq\label{e4}
T_{tt}=\la(\dot{\rho}\gamma + \dot{\beta}).
\feq
(The constant of proportionality has been chosen  to ensure that
$T_{tt}$ has the dimensions of a mass squared). In fact, from
Eqns.~(\ref{d5}), (\ref{d5a}) and (\ref{d7}) it is evident that
$T_{tt}$ plays the same role as the holomorphic $T_{++}$
(and antiholomorphic $T_{--}$) stress-energy tensor play in 2D
CFT. A more compelling argument leading to Eq.~(\ref{e4}) relies on the
identification of $T_{tt}$ as the boundary value for the
stress-energy tensor of a 2D CFT.

To do this, we choose the conformal gauge
\eq\lb{g1}
ds^{2}= -e^{2\om}dx^{+}dx^{-}.
\feq
Then the action (\ref{JT}) takes the Liouville-like form
\eq
I = -\int d^2x \left(\partial_{+}\om \partial_{-}\eta+
\partial_{-}\om \partial_{+}\eta - {1\over 2}\la^{2}\eta
e^{2\om}\right).
\label{JT1}
\feq
The action must be complemented by the constraints (the equations of
motion for the missing components of the metric)
\eq\lb{g2}
T_{\pm\pm}= \partial_{\pm}^{2}\eta -2 \partial_{\pm}\eta
\partial_{\pm}\om=0,
\feq
where $T_{\pm\pm}$ denote the components of the stress-energy tensor.

For $\eta\to\infty$ the potential term in the action (\ref{JT1}) goes to
zero and the model becomes an exact 2D CFT \cite{CC}.
In fact, defining the new fields $X,Y$,

\eq\lb{g3}
\om=X-Y,\qquad \eta=X+Y,
\feq
the action and the constraints become
\eqn
I &=& -2\int d^2x \left(\partial_{+}X \partial_{-}X -
\partial_{+}Y \partial_{-}Y \right), \\
\label{JT2}
T_{\pm\pm} &=& \partial_{\pm}^{2}X +  \partial_{\pm}^{2}Y
-2 \partial_{\pm}X\partial_{\pm}X+2\partial_{\pm}Y\partial_{\pm}Y=0.
\label{g4}
\feqn
Taking $\eta\to\infty$ we reach the boundary of AdS$_{2}$,
which in light-cone coordinates is located at
$x^{+}=x^{-}$.
One can show that $T_{++}|_{boundary}=T_{--}|_{boundary}=T_{tt}$,
with $T_{tt}$ given by Eq.~(\ref{e4}).

Let us now show that $T_{tt}$ generates the diff$_1$ group.
Introducing the charges
\eq\label{e4a}
\hat J=\int \epsilon T_{tt},
\feq
and using the transformation law (\ref{d7}), one has
\eq\label{e5}
\delta_{\epsilon}T_{tt}=[\hat J, T_{tt}] =
\epsilon \dot{T_{tt}} + 2\dot{\epsilon}T_{tt}.
\feq
Expanding in Fourier modes,
\eq\label{e6}
T_{tt}=\sum L_{m} e^{-i m \lambda t},\qquad \epsilon(t)= \sum a_{m}
e^{-i m \lambda t},
\feq
and using Eq.~(\ref{e5}), one finds that $L_{m}$ generates a Virasoro
algebra,
\eq\label{e6a}
[L_{m}, L_{m}]= (m-n) L_{m+n}.
\feq

Eq. (\ref{e4}) does not give the most general form of the CFT$_{1}$
stress-energy tensor. We have the freedom to add to it a constant term
(which we choose proportional to the black hole mass $M$) and an  
improvement term,
\eq\label{f1}
T_{tt}|_{impr} = T_{tt} +\la M+ b\ddot\rho,
\feq
where $b$ is an arbitrary constant.
The improvement  term is a total derivative and does not affect the 
status
of $T_{tt}$ as generator of the diff$_{1}$ group.
Though the constant $b$ is at the CFT level undetermined,
it can be computed using the underlying gravitational dynamics.
Comparing (\ref{f1}) with the expression for
$T_{tt}$ found in Ref.~\cite{CM00}, one gets $b= - 2\eta_{0}$.

We can now calculate the central charge $C$ associated with the
Virasoro algebra (\ref{e6a}).
A naive computation will
give the value $C=24\eta_{0}$ found in Ref.~\cite{CM00}.
This can be shown explicitly using
the general transformation law of a CFT$_{1}$ stress-energy tensor,

\eq\label{f3}
\delta T_{tt}=\epsilon \dot  T_{tt} + 2\dot \epsilon  T_{tt}+
{C\over 12} \stackrel\dots \epsilon.
\feq
Using Eqns.~(\ref{d6}) and (\ref{d7}), one can easily show that
$T_{tt}$ given by Eq. (\ref{f1}) follows the transformation law
(\ref{f3}), with $C=24\eta_0$.
This value of the central charge, once inserted in the Cardy formula,
$S=2\pi\sqrt{CL _{0} \over 6}$ \cite{cardy}, produces a black hole entropy,
which differs by a $\sqrt 2$
factor from the thermodynamical value (\ref{entropy}).

However in this computation (and in those of Refs.~\cite{CM99},
\cite{CM00} as well), one only considers the contribution of the
$r\to\infty$ boundary of AdS$_2$. The black hole solution (\ref{bh})
has also an
inner boundary \cite{CM95,CM99}
(located at $r=0$ for the ground state or at the horizon for the
generic black hole), which can give a contribution  to the central charge.
That this inner boundary can be crucial for understanding the black
hole entropy has been shown in \cite{CCV}.

There is a simple way to compute this contribution.
We need to change the coordinates, from the Schwarzschild $(r,t)$
frame used in Eq.~(\ref{bh}) to the conformal frame, where the vacuum and
black hole solutions have, respectively, the form \cite{CM95},

\eq\lb{f4}
ds^{2}={1\over \la^2x^2} (-dt^2 +dx^2),
\feq

\eq\label{f5}
ds^2={a^{2}\over\sinh^2(a\la\si)}(-d\tau ^2+d\si^2).
\feq
The key point is that in the conformal frame the inner boundary (and the
horizon) is pushed to $x=\infty$, whereas the the timelike $r=\infty$ boundary
of AdS$_2$ is now located at $x=0$. The information about the existence of the
inner boundary is now encoded
in the coordinate transformation
\eq\label{f6}
t={1\over a \la}e^{a\la\t}\cosh(a\la\si)\,,\qquad
x= {1\over a \la} e^{a\la \t}\sinh(a\la\si)\,,
\feq
that maps the vacuum (\ref{f4}) into the black hole solution (\ref{f5})
\cite{CM95}.

Evaluating these transformations on the $x=\si=0$ boundary, where our
one- dimensional conformal field theory lives, we find

\eq\label{f7}
t={1\over a \la}e^{a\la\t}.
\feq

Thus the vacuum and black hole solutions correspond to different
time variables on the boundary. Moreover, for $-\infty<\t<\infty$,
$0<t<\infty$,
part of the "history" seen by the vacuum observer cannot be seen
by the black hole observer. It follows that there will be a term
in the entropy describing the entanglement of states, which has the
form of a contribution $C_{ent}$ to the central charge.
This contribution can be calculated using a method similar to that
employed in Ref.~\cite{CC}
($C_{ent}$ is interpreted as a Casimir energy).

The transformation law of the
stress-energy tensor under a general change of coordinates $t=t(\t)$
is given by the transformation (\ref{f3}) in its finite form,

\eq\lb{f8}
T_{\t\t}=\left ({dt\over d\t}\right )^{2}T_{tt}- {C_{ent}\over 12}
\left ({dt\over
d\t}\right)^{2}\{\t,t\}\,,
\feq
where $\{\t,t\}$ is the Schwarzian derivative.
Applied to the transformation (\ref{f7}), Eq.~(\ref{f8}) gives
\eq\lb{f8a}
T_{\t\t}= (a\la t)^{2}T_{tt}- {C_{ent}\over 24} a^{2}\la^{2} \,.
\feq

From Eq. (\ref{f1}) it follows that by fixing the diffeomorphisms
invariance, on-shell we can always have
$T_{\t\t}= \la M$. In fact, on-shell the term proportional to the
constraints is zero and we can always choose $\rho=const$.
Because $T_{tt}$ in Eq.~(\ref{f8a}) refers to the vacuum, we have
$T_{tt}=0$, and (\ref{f8a}) becomes

\eq\lb{f8b}
\la M= - {C_{ent}\over 24} a^{2}\la^{2}¥\,.
\feq

The coordinate transformation (\ref{f7}) maps the ground state
into the black hole with mass
$M=a^{2}\eta_{0}\la/2$ \cite{CM95}, which inserted into Eq.~(\ref{f8b})
gives

\eq\lb{f9}
C_{ent}=-12 \eta_{0}\,.
\feq
Notice that the entanglement contribution is negative, yielding a total
central charge

\eq\lb{f10}
C_{tot}=C+C_{ent}= 12 \eta_{0}.
\feq
Using this value of the central charge in Cardy's formula, one finds perfect
agreement with the thermodynamical entropy of the 2D black hole
(\ref{entropy}).

Let us now discuss the physical interpretation of the dynamical
system (\ref{dffeq}) and its relationship with 2D dilaton gravity.
The equation of motion  (\ref{dffeq}) describes a mechanical system
coupled to an external source. Alternatively, one can think of
$\gamma$ as a time-dependent coupling constant,
appearing in the harmonic oscillator potential.

Because $\gamma$ is arbitrary (the only constraint on it is the
transformation law (\ref{d6})), the dynamical system is
essentially non-deterministic. Moreover, because the external source
is time-dependent, the energy is not conserved and we have a
time-dependent Hamiltonian, whose evolution is not completely fixed
by the dynamics of the system.

Strictly speaking we have to deal with
an ensemble of Hamiltonians. One has thus a strong analogy with disordered
systems in statistical mechanics. The main difference is that, whereas in
the case of disordered systems we have a probability distribution for
the couplings, in our case they are arbitrary functions, for which we
only give the transformation law under the conformal group.
Moreover, in our case $\gamma$ is a smooth function of $t$.

From the point of view of the 2D gravitational theory the meaning of
the source $\gamma$ is clear.
The fields $\beta$ and $\gamma$ describe deformations of the boundary
of AdS$_2$, generated by 2D bulk diffeomorphisms, whereas the field $\rho$
describes deformations of the dilaton.
Thus, the function $\gamma$  encodes the information about the gauge
symmetry of the 2D gravity theory. The non-deterministic nature of the
dynamical system (\ref{dffeq}) is a consequence of the gauge freedom
of the gravitational dynamics. This indicates
an interesting relationship between
gauge symmetries and non-deterministic dynamical systems.

In Ref.~\cite{CM00}, it was pointed out that
the non constant value of the dilaton ($\rho\neq 0$ in terms of
boundary fields) breaks the $SL(2,R)$ isometry
group of AdS$_2$, and that the origin of the central charge (and hence
of the black hole entropy) can be traced back to this breaking.
Different values of $\rho$ represent different vacua of the 2D
gravity theory which break $SL(2,R)$. Indeed the conformal
transformations (\ref{d6}), (which are the boundary counterpart of the
2D diffeomorphisms) map all these vacua one into the other. Moreover, because
the energy (\ref{energy}) is invariant (on-shell) under conformal
transformations, all these vacua are degenerate in energy.

Now, the crucial point is that there is a one-to-one correspondence
between these vacua and the solutions of the dynamical system
(\ref{dffeq}). Note that also the (IR-regularized) DFF
model (\ref{1daction}) breaks the $SL(2,R)$ symmetry of the
original model if we take the source $\gamma$
constant. Introducing a time-dependent external source,
transforming  as conformal field of weight 2, we reinstate
the full conformal symmetry. From the point of view of 2D gravity
we are considering different $\rho$-dependent vacua.

The above considerations
indicate a natural way to explain statistically the
entropy of 2D black holes.
This entropy can be
interpreted in terms of the degeneracy of the $\rho$-vacua.
The energy (\ref{energy}) is invariant under conformal
transformations, so that we can calculate the entropy by counting the
independent excitations in the configuration space of vacua.

From the point of view of the dynamical system this degeneracy is
encoded in the external source $\gamma$.
The quadratic mass-temperature and mass-entropy dependence \cite{CC},
which is typical of a 2D CFT and which, in principle, could be used
to rule out the duality of AdS$_2$ gravity with a conformal quantum
mechanics, is presumably related to the fact that
the conformally invariant mechanical system in question
is not a usual mechanical system but a DFF model coupled to an external
source.

\end{document}